\documentstyle[epsf]{article}

\newcommand{\be}{\begin{equation}}
\newcommand{\ee}{\end{equation}}
\newcommand{\ba}{\begin{eqnarray}}
\newcommand{\ea}{\end{eqnarray}}
\newcommand{\no}{\nonumber\\}

\newcommand{\grts}{\raise.3ex\hbox{$>$\kern-.75em\lower1ex\hbox{$\sim$}}}
\newcommand{\lets}{\raise.3ex\hbox{$<$\kern-.75em\lower1ex\hbox{$\sim$}}}

\title{Real CP violation \\ in a simple extension of the standard model}

\author{L.\ Lavoura \\
\small Universidade T\'ecnica de Lisboa \\
\small Centro de F\'\i sica das Interac\c c\~oes Fundamentais \\
\small Instituto Superior T\'ecnico, 1049-001 Lisboa, Portugal}

\date{7 September 1999}

\begin{document}
\maketitle

\begin{abstract}
I present a simple three-Higgs-doublet extension of the standard model
in which real CP violation takes place.
The strong CP problem is attenuated by this model.
\end{abstract}

\section{Introduction}

The concept of real CP violation has recently been introduced
by Masiero and Yanagida \cite{masiero}.
It stands for spontaneous CP violation (SCPV)
which occurs in spite of the vacuum expectation values (VEVs) being real.
In a model with real CP violation,
complex numbers are present in the Yukawa couplings from the very beginning;
however,
unless the (real) VEVs break a certain CP symmetry,
those complexities cancel among themselves,
and the whole theory is CP conserving.
As a matter of fact,
a model of real CP violation had already been suggested
by myself \cite{lavoura} before Masiero and Yanagida's paper.

The possibility that CP might be spontaneously broken
by real VEVs should not come as a surprise.
This may simply happen because the basic Lagrangian
(before spontaneous symmetry breaking)
is invariant under a non-trivial CP transformation,
as I shall now explain with the help of a simple example.
Suppose that there are two non-Hermitian scalar fields,
$S_1$ and $S_2$,
which transform under CP into each other's Hermitian conjugate:
\be
S_1 \stackrel{\rm CP}{\rightarrow} S_2^\dagger,\
S_2 \stackrel{\rm CP}{\rightarrow} S_1^\dagger.
\label{CP example}
\ee
If the VEVs of $S_1$ and $S_2$ have different modulus,
then the CP transformation of Eq.~(\ref{CP example}) is broken.
This happens independently of the phases of the VEVs.
On the other hand,
we may define
\be
S_\pm \equiv \frac{S_1 + S_2 \pm i \left( S_1 - S_2 \right)}{2}.
\label{re-definition}
\ee
Then,
the CP transformation of Eq.~(\ref{CP example})
looks like the usual CP transformation:
\be
S_+ \stackrel{\rm CP}{\rightarrow} S_+^\dagger,\
S_- \stackrel{\rm CP}{\rightarrow} S_-^\dagger.
\label{CP new version}
\ee
When the VEVs of $S_1$ and $S_2$ are real and distinct,
the VEVs of $S_\pm$ are complex.
This we would describe as being just the usual form of SCPV.

Thus,
the situation which in the basis $\left\{ S_1, S_2 \right\}$
would be described as real CP violation,
looks in the basis $\left\{ S_+, S_- \right\}$ like the usual SCPV,
with complex VEVs.
One may then ask oneself whether the concept of real CP violation
is not altogether spurious.
But,
using $S_+$ and $S_-$ as the basic fields,
instead of using $S_1$ and $S_2$,
may be inappropriate,
in particular if there is in the theory a symmetry,
besides CP,
under which $S_1$ and $S_2$ transform as singlets,
while $S_+$ and $S_-$ mix.
For instance,
there may be an extra symmetry under which
\be
S_1 \rightarrow e^{2 i \pi / 3} S_1,\
S_2 \rightarrow e^{- 2 i \pi / 3} S_2.
\label{extra symmetry}
\ee
Clearly,
$S_+$ and $S_-$ mix under the transformation in Eq.~(\ref{extra symmetry}).
Hence,
they form an inadequate basis to study the theory,
and it is more fruitful to use the basis $\left\{ S_1, S_2 \right\}$.
It is then appropriate to say that there is real CP violation.
On the other hand,
if no symmetry like the one in Eq.~(\ref{extra symmetry}) exists,
then the concept of real CP violation has no distinctive meaning,
as it is just as legitimate to study the theory
in the basis $\left\{ S_+, S_- \right\}$
as in the basis $\left\{ S_1, S_2 \right\}$.

Alternatively,
one may argue that,
if a model with real CP violation is able to attain
some original and useful achievement,
any result which would not in general obtain,
then real CP violation is useful and worth taking seriously.
In their paper \cite{masiero},
Masiero and Yanagida have suggested that real CP violation
might provide an avenue to solving the strong CP problem;
as a matter of fact,
a realization of this prophecy had already been provided
in my paper \cite{lavoura}.
An alternative application of real CP violation,
and of the associated non-trivial CP symmetry,
might be to constrain the fermion mass matrices,
leading to the prediction of some relationships
among the fermion masses and mixing angles.
This would be analogous to what was done by Ecker and collaborators
\cite{ecker} in the context of left--right-symmetric models.

Both the models of real CP violation
presented by Masiero and Yanagida \cite{masiero} and by myself \cite{lavoura}
have many fields beyond the standard-model ones.
Besides,
those models also include extra non-Abelian symmetry groups.
This might lead one to believe that real CP violation is an exotic effect,
which can occur or be useful only in the context of complicated models.
The main purpose of this paper to show that this is not so.
Specifically,
I construct here a simple extension of the standard model,
with no extra fermions,
no extra gauge groups,
and no extra non-Abelian symmetries,
in which real CP violation occurs.

The second goal of this paper is
to call attention to a potential advantage of real CP violation.
Namely,
one is able to get spontaneous breaking of CP
without thereby introducing CP violation
in the mixings and self-interactions of the scalar fields.
The third aim of this paper is to use that property
to attenuate the strong CP problem of the standard model.
Indeed,
once CP violation is removed from the scalar sector,
then the possibility arises of eliminating the generation
of an unacceptably large $\theta_{\rm QFD}$
through the one-loop diagram of Fig.~1 \cite{georgi}.
\begin{figure}[tb]
\begin{center}
\leavevmode \epsfxsize=120mm \epsfbox{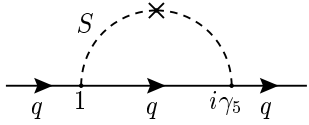}
\end{center}
\vspace*{-12mm}
\caption{One-loop diagram which may generate a large $\theta_{\rm QFD}$.}
\label{Georgi's diagram}
\vspace*{-3mm}
\end{figure}
When scalar--pseudoscalar mixing is absent that diagram does not exist,
and a non-zero $\theta_{\rm QFD}$ may arise only at two-loop level,
it being then expected to be of order $10^{-8}$.
This would milden the strong CP problem.
Still,
I cannot claim that
the specific model in the following section does solve that problem.

\section{The model}

The model that I want to put forward in order to materialize my claims
has three Higgs doublets.
The gauge group is SU(2)$\otimes$U(1)
and the fermion spectrum is the usual one.
In particular,
there are three left-handed-quark doublets
$q_{La} = \left( p_{La}, n_{La} \right)^T$,
three right-handed charge-$2/3$ (up-type) quarks $p_{Ra}$,
and three right-handed charge-$-1/3$ (down-type) quarks $n_{Ra}$.
(The index $a$ runs from $1$ to $3$.)
There are three scalar doublets
\be
\phi_a = e^{i \theta_a} \left( \begin{array}{c} \varphi_a^+ \\*[1mm]
v_a + {\displaystyle \frac{\rho_a + i \eta_a}{\sqrt{2}}} \end{array} \right),
\ee
where $v_a e^{i \theta_a}$ are the VEVs,
with $v_a$ real and positive by definition.
The fields $\rho_a$ and $\eta_a$ are Hermitian.
As usual,
I denote
\be
\tilde \phi_a \equiv i \tau_2 {\phi_a^\dagger}^T = e^{- i \theta_a}
\left( \begin{array}{c} v_a + {\displaystyle
\frac{\rho_a - i \eta_a}{\sqrt{2}}} \\*[1mm] - \varphi_a^- \end{array} \right).
\ee

There is in the model a discrete symmetry $D$ under which
\be
\phi_1 \stackrel{D}{\rightarrow} i \phi_1,\
\phi_2 \stackrel{D}{\rightarrow} -i \phi_2,\
q_{L1} \stackrel{D}{\rightarrow} i q_{L1},\
q_{L2} \stackrel{D}{\rightarrow} -i q_{L2},
\label{simetria D}
\ee
and all other fields remain invariant.

The symmetry CP transforms the right-handed-quark fields in the usual way,
\be
p_{Ra} \stackrel{\rm CP}{\rightarrow} \gamma^0 C \overline{p_{Ra}}^T,\
n_{Ra} \stackrel{\rm CP}{\rightarrow} \gamma^0 C \overline{n_{Ra}}^T,
\label{CP de direita}
\ee
but it interchanges the indices $1$ and $2$
in the transformation of both the scalar doublets
and the left-handed-quark doublets:
\be
\phi_1 \stackrel{\rm CP}{\rightarrow} {\phi_2^\dagger}^T,\
\phi_2 \stackrel{\rm CP}{\rightarrow} {\phi_1^\dagger}^T,\
\phi_3 \stackrel{\rm CP}{\rightarrow} \pm {\phi_3^\dagger}^T,
\label{CP scalars}
\ee
\be
q_{L1} \stackrel{\rm CP}{\rightarrow} \gamma^0 C \overline{q_{L2}}^T,\
q_{L2} \stackrel{\rm CP}{\rightarrow} \gamma^0 C \overline{q_{L1}}^T,\
q_{L3} \stackrel{\rm CP}{\rightarrow} \pm \gamma^0 C \overline{q_{L3}}^T.
\label{CP de esquerda}
\ee
The fields in the left-hand side
of Eqs.~(\ref{CP de direita})--(\ref{CP de esquerda})
are meant to be at the space--time point $\left( t, \vec r \right)$,
while the fields in the right-hand side
are at the space--time point $\left( t, - \vec r \right)$.
The $\pm$ sign in Eqs.~(\ref{CP scalars}) and (\ref{CP de esquerda})
means that there are as a matter of fact two different CP symmetries,
both of which leave the Lagrangian of the model invariant.

Notice that the transformations $D$ and CP commute.

It follows from Eq.~(\ref{CP scalars}) that
CP will be preserved by the vacuum if
\be
v_1 e^{i \theta_1} = v_2 e^{- i \theta_2},\
v_3 e^{i \theta_3} = \pm v_3 e^{- i \theta_3}.
\label{no SCPV, 1}
\ee
The phases $\theta_a$ are gauge-dependent,
and therefore the conditions in Eq.~(\ref{no SCPV, 1})
are not gauge-invariant.
The invariant conditions for the absence of SCPV are
\be
v_1 = v_2,\ e^{i \left( 2 \theta_3 - \theta_1 - \theta_2 \right)} = \pm 1.
\label{no SCPV, 2}
\ee
One sees that SCPV may be achieved through $v_1 \neq v_2$,
quite independently of the phases of the VEVs.
This situation embodies what I call real CP violation.

As a consequence of the symmetries $D$ and CP,
the scalar potential $V$ has only two terms
which ``see'' the relative phases of the doublets.
Indeed,
$V = V_I + V_S$,
where
\ba
V_I &=&
\mu_1 \left( \phi_1^\dagger \phi_1 + \phi_2^\dagger \phi_2 \right)
+ \mu_2 \phi_3^\dagger \phi_3
+ \lambda_1 \left[ \left( \phi_1^\dagger \phi_1 \right)^2
+ \left( \phi_2^\dagger \phi_2 \right)^2 \right]
+ \lambda_2 \left( \phi_3^\dagger \phi_3 \right)^2
\no & &
+ \lambda_3 \left( \phi_1^\dagger \phi_1 \right)
\left( \phi_2^\dagger \phi_2 \right)
+ \lambda_4 \left[ \left( \phi_1^\dagger \phi_1 \right)
+ \left( \phi_2^\dagger \phi_2 \right) \right]
\left( \phi_3^\dagger \phi_3 \right)
\no & &
+ \lambda_5 \left( \phi_1^\dagger \phi_2 \right)
\left( \phi_2^\dagger \phi_1 \right)
+ \lambda_6 \left[ \left( \phi_1^\dagger \phi_3 \right)
\left( \phi_3^\dagger \phi_1 \right) + \left( \phi_2^\dagger \phi_3 \right)
\left( \phi_3^\dagger \phi_2 \right) \right]
\ea
and
\ba
V_S &=& \lambda_7 \left[
\left( \phi_3^\dagger \phi_1 \right) \left( \phi_3^\dagger \phi_2 \right)
+ \left( \phi_1^\dagger \phi_3 \right) \left( \phi_2^\dagger \phi_3 \right)
\right]
\no & &
+ \lambda_8 \left[ e^{i \chi} \left( \phi_1^\dagger \phi_2 \right)^2
+ e^{- i \chi} \left( \phi_2^\dagger \phi_1 \right)^2 \right].
\ea
$V_I$ is that part of $V$
which is insensitive to the overall phases of the doublets.
The coefficients $\mu_1$,
$\mu_2$,
and $\lambda_{1\mbox{--}8}$ are real by Hermiticity.
Notice that our particular CP symmetry
allows an unconstrained phase $\chi$ to appear in the potential.
As there are two gauge-invariant phases among the VEVs
and two terms in the potential which ``see'' those phases,
no undesirable Goldstone bosons arise,
and the phases of the VEVs adjust in such a way that
\be
e^{i \left( 2 \theta_3 - \theta_1 - \theta_2 \right)} = \left( -1 \right)^a,\
e^{i \left( 2 \theta_2 - 2 \theta_1 + \chi \right)} = \left( -1 \right)^b,
\label{vacuum phases}
\ee
where $a$ and $b$ are either $0$ or $1$,
and
\be
\lambda_7 \left( -1 \right)^a = - \left| \lambda_7 \right|,\
\lambda_8 \left( -1 \right)^b = - \left| \lambda_8 \right|.
\label{signs of the VEVs}
\ee

The first condition in Eq.~(\ref{vacuum phases})
means that SCPV cannot be achieved through violation
of the second relation in Eq.~(\ref{no SCPV, 2}).
Still,
$v_1 \neq v_2$ is possible because of the presence
of the coupling $\lambda_7$ in the scalar potential.
Indeed,
the stability conditions for the VEVs,
assuming $v_1 \neq v_2$,
are
\ba
\mu_1 &=& - 2 \lambda_1 \left( v_1^2 + v_2^2 \right)
- \left( \lambda_4 + \lambda_6 \right) v_3^2, \no
\mu_2 &=& - 2 \lambda_2 v_3^2
- \left( \lambda_4 + \lambda_6 \right) \left( v_1^2 + v_2^2 \right)
+ 2 \left| \lambda_7 \right| v_1 v_2, \\
\left| \lambda_7 \right| &=& \left( - 2 \lambda_1 + \lambda_3 + \lambda_5
- 2 \left| \lambda_8 \right| \right) v_1 v_2 /v_3^2.
\nonumber
\ea

Thus,
in this model CP is spontaneously broken through $v_1 \neq v_2$.
The VEVs are not real,
because of the phase $\chi$ coming from $V_S$,
but their phases are immaterial for SCPV,
as they merely adjust in order to offset
the phases in the scalar potential---see
Eqs.~(\ref{vacuum phases}) and (\ref{signs of the VEVs}).
As a remarkable consequence,
even when $v_1 \neq v_2$,
CP violation remains absent from the self-interactions of the scalars.
In particular,
there are two physical charged scalars $H_1^+$ and $H_2^+$,
\be
\left( \begin{array}{c} G^+ \\ H_1^+ \\ H_2^+ \end{array} \right)
= T_H \left( \begin{array}{c} \varphi_1^+ \\ \varphi_2^+ \\ \varphi_3^+
\end{array} \right),
\label{charged-scalar mixing}
\ee
and the $3 \times 3$ matrix $T_H$ is orthogonal,
{\it i.e.},
{\em real}.
The field $G^+$ is the Goldstone boson which is absorbed
by the longitudinal component of the $W^+$.
There are two physical neutral pseudoscalars $A_1$ and $A_2$,
\be
\left( \begin{array}{c} G^0 \\ A_1 \\ A_2 \end{array} \right)
= T_A \left( \begin{array}{c} \eta_1 \\ \eta_2 \\ \eta_3
\end{array} \right),
\label{pseudoscalar mixing}
\ee
where $G^0$ is the Goldstone boson which is absorbed by the $Z^0$.
The first rows of $T_H$ and of $T_A$
are given by $\left( v_1, v_2, v_3 \right) / v$,
where $v = \sqrt{v_1^2 + v_2^2 + v_3^2} = 174\, {\rm GeV}$.
Finally,
there are three physical neutral scalars $N_a$,
\be
\left( \begin{array}{c} N_1 \\ N_2 \\ N_3 \end{array} \right)
= T_N \left( \begin{array}{c} \rho_1 \\ \rho_2 \\ \rho_3
\end{array} \right).
\label{scalar mixing}
\ee
The matrices $T_A$ and $T_N$ are orthogonal.
The important point is that
{\em the fields $\rho$ do not mix with the fields $\eta$}.
This,
and the reality of $T_H$,
are consequences of the CP invariance of the scalar potential,
which is retained even after SCPV has been achieved through $v_1 \neq v_2$.

The Yukawa Lagrangian of the quarks is
\ba
{\cal L}_{\rm Y}^{\rm (q)} &=& - \left( \overline{q_{L1}} \phi_1 \Gamma_1 +
\overline{q_{L2}} \phi_2 \Gamma_1^\ast + \overline{q_{L3}} \phi_3 \Gamma_2
\right) n_R
\no & & - \left( \overline{q_{L1}} \tilde \phi_2 \Delta_1^\ast +
\overline{q_{L2}} \tilde \phi_1 \Delta_1 +
\overline{q_{L3}} \tilde \phi_3 \Delta_2 \right) p_R
+ {\rm H.c.}.
\label{Yukawas 1}
\ea
$\Gamma_1$,
$\Gamma_2$,
$\Delta_1$,
and $\Delta_2$ are $1 \times 3$ row matrices.
Because of CP symmetry,
$\Gamma_2$ and $\Delta_2$ are real;
on the other hand,
$\Gamma_1$ and $\Delta_1$ are in general complex.
The mass matrices of the quarks are of the form
\be
M_n = \left( \begin{array}{c} v_1 e^{i \theta_1} \Gamma_1 \\
v_2 e^{i \theta_2} \Gamma_1^\ast \\ v_3 e^{i \theta_3} \Gamma_2
\end{array} \right),\
M_p = \left( \begin{array}{c} v_2 e^{- i \theta_2} \Delta_1^\ast \\
v_1 e^{- i \theta_1} \Delta_1 \\ v_3 e^{- i \theta_3} \Delta_2
\end{array} \right).
\label{mass matrices}
\ee
The matrices $M_n$ and $M_p$ are bi-diagonalized
by unitary matrices $U^n_{L,R}$ and $U^p_{L,R}$:
\be
{U_L^n}^\dagger M_n U^n_R = M_d = {\rm diag} \left( m_d, m_s, m_b \right),\
{U_L^p}^\dagger M_p U^p_R = M_u = {\rm diag} \left( m_u, m_c, m_t \right).
\label{bi-diagonalization}
\ee
The relationship between the quarks in the weak basis and in the mass basis is
\be
n_L = U_L^n d_L,\
n_R = U_R^n d_R,\
p_L = U_L^p u_L,\
p_R = U_R^p u_R.
\label{unitary transformation}
\ee
The Cabibbo--Kobayashi--Maskawa (CKM) matrix is
\be
V = {U_L^p}^\dagger U_L^n.
\ee
Because of the form of $M_n$ and $M_p$ in Eqs.~(\ref{mass matrices}),
and in particular as $\Gamma_2$ and $\Delta_2$ are real,
$V$ is intrinsically complex---{\it i.e.},
$J \equiv {\rm Im} \left( V_{us} V_{cb} V_{ub}^\ast V_{cs}^\ast \right)
\neq 0$---if and only if $v_1 \neq v_2$.
Indeed,
when $\sin \left( 2 \theta_3 - \theta_1 - \theta_2 \right) = 0$
one finds that
\be
\det \left( M_p M_p^\dagger M_n M_n^\dagger
- M_n M_n^\dagger M_p M_p^\dagger \right) \propto v_2^2 - v_1^2.
\label{trace}
\ee
As is well known \cite{livro},
$J$ is proportional to the determinant in Eq.~(\ref{trace}).

Thus,
in this model the spontaneous breaking of CP
does not lead to CP violation in the scalar sector,
but it does lead to a complex CKM matrix.
The appearance of CP violation in this model is completely independent
of the values of the phases of the VEVs.
Indeed,
even though $\theta_2 - \theta_1$ and $\theta_3 - \theta_1$
are in general non-zero---because of the arbitrary phase $\chi$ in $V_S$,
see Eqs.~(\ref{vacuum phases})---the appearance of CP violation
hinges on $v_1 \neq v_2$,
and not on the values of $\theta_2 - \theta_1$ and $\theta_3 - \theta_1$.
It is thus appropriate to describe this model
as displaying real CP violation,
even though the VEVs may have non-zero relative phases.

\section{Consequences for strong CP violation}

As is well known,
non-perturbative effects in Quantum Chromodynamics (QCD)
may lead to P and CP violation,
characterized by a parameter $\theta = \theta_{\rm QCD} + \theta_{\rm QFD}$,
in hadronic processes.
The first contribution to $\theta$,
namely $\theta_{\rm QCD}$,
characterizes P and CP violation in the vacuum of QCD.
This contribution to $\theta$ may be assumed to vanish in a model,
like the present one,
whose Lagrangian conserves CP.
On the other hand,
\be
\theta_{\rm QFD} = \arg\, \det \left( M_n M_p \right)
\ee
does not in general vanish.
This is because the quark mass matrices must be complex---lest
the CKM matrix be real---and there is in general no reason
why their determinant should be real.

From the experimental upper bound on the electric dipole moment of the neutron
one finds that $\theta\, \lets\, 10^{- \left( 9\mbox{--}10 \right)}$
\cite{ritz}.
The presence of such a small number in QCD
constitutes the strong CP problem.\footnote{The situation
$\theta = \pi$ is equivalent to $\theta = 0$,
in that it also corresponds to the absence of P and CP violation
by the strong interaction.}

In the present model,
$\theta_{\rm QCD}$ is zero because CP is a symmetry of the Lagrangian.
As for $\theta_{\rm QFD}$,
one gathers from the mass matrices in Eq.~(\ref{mass matrices}) that
\ba
\arg\, \det M_n &=& \theta_1 + \theta_2 + \theta_3 + \frac{\pi}{2}
\left( \mbox{mod}\ \pi \right),
\no
\arg\, \det M_p &=& - \theta_1 - \theta_2 - \theta_3 - \frac{\pi}{2}
\left( \mbox{mod}\ \pi \right),
\ea
and therefore $\theta_{\rm QFD} = 0$ at tree level.

A non-zero $\theta_{\rm QFD}$ may however be generated at loop level
\cite{ellis}.
Let us denote the loop contributions to the mass matrices
of the down-type and up-type quarks,
as computed in the basis of the tree-level physical quarks $d$ and $u$,
by $\Sigma_d$ and $\Sigma_u$,
respectively.
Thus,
the quark mass Lagrangian at loop level is
\be
{\cal L}_{\rm mass}^{\rm (q)} =
- \overline{d_L} \left( M_d + \Sigma_d \right) d_R
- \overline{u_L} \left( M_u + \Sigma_u \right) u_R
+ {\rm H.c.}.
\ee
Then,
\ba
\theta_{\rm QFD} &=& \arg\, \det \left( M_d + \Sigma_d \right)
+ \arg\, \det \left( M_u + \Sigma_u \right)
\no &=& \arg\, \det\, M_d + \arg\, \det \left( 1 + \Sigma_d M_d^{-1} \right)
+ \arg\, \det\, M_u + \arg\, \det \left( 1 + \Sigma_u M_u^{-1} \right)
\no &=& \arg\, \det \left( 1 + \Sigma_d M_d^{-1} \right)
+ \arg\, \det \left( 1 + \Sigma_u M_u^{-1} \right)
\no &=& \mbox{Im tr}\, \ln \left( 1 + \Sigma_d M_d^{-1} \right)
+ \mbox{Im tr}\, \ln \left( 1 + \Sigma_u M_u^{-1} \right).
\label{theta-loop}
\ea
I have used the fact that $M_d$ and $M_u$ are real matrices.
The expression in Eq.~(\ref{theta-loop}) may be computed
with the help of the identity,
valid for small $C$,
\be
\ln \left( 1 + C \right) = C - \frac{C^2}{2} + \frac{C^3}{3} - \cdots.
\ee

The dangerous contributions to the one-loop quark self-energies
are the ones from diagrams with either charged or neutral scalars in the loop
\cite{georgi},
as in Fig.~1.
Let us therefore compute the Yukawa interactions of the model
in the physical-quark basis.
I first write the unitary matrices $U_L^n$ and $U_L^p$ as
\be
U_L^n = \left( \begin{array}{c} N_1 \\ N_2 \\ N_3 \end{array} \right),\
U_L^p = \left( \begin{array}{c} P_1 \\ P_2 \\ P_3 \end{array} \right),
\ee
where the $N_a$ and the $P_a$ are $1 \times 3$ row matrices.
Let us define $A_a \equiv P_a^\dagger N_a$.
Clearly,
the CKM matrix
\be
V = {U_L^p}^\dagger U_L^n = A_1 + A_2 + A_3.
\ee
The unitarity of $U_L^n$ and $U_L^p$ implies
\ba
A_a A_b^\dagger &=& \delta_{ab} P_a^\dagger P_a,
\no
A_a^\dagger A_b &=& \delta_{ab} N_a^\dagger N_a.
\label{A A = delta}
\ea
Moreover,
\be
A_1 A_1^\dagger + A_2 A_2^\dagger + A_3 A_3^\dagger =
A_1^\dagger A_1 + A_2^\dagger A_2 + A_3^\dagger A_3 = 1_{3 \times 3}.
\ee
The quark Yukawa Lagrangian in Eq.~(\ref{Yukawas 1}) may,
when one takes into account Eqs.~(\ref{bi-diagonalization})
and (\ref{unitary transformation}),
be written as
\ba
{\cal L}_{\rm Y}^{\rm (q)} &=& - \overline{d_L} M_d d_R
- \overline{u_L} M_u u_R
\no & &
- \overline{d_L} \left(
\frac{\rho_1 + i \eta_1}{\sqrt{2} v_1} A_1^\dagger A_1
+ \frac{\rho_2 + i \eta_2}{\sqrt{2} v_2} A_2^\dagger A_2
+ \frac{\rho_3 + i \eta_3}{\sqrt{2} v_3} A_3^\dagger A_3 \right) M_d d_R
\no & &
- \overline{u_L} \left(
\frac{\rho_2 - i \eta_2}{\sqrt{2} v_2} A_1 A_1^\dagger
+ \frac{\rho_1 - i \eta_1}{\sqrt{2} v_1} A_2 A_2^\dagger
+ \frac{\rho_3 - i \eta_3}{\sqrt{2} v_3} A_3 A_3^\dagger \right) M_u u_R
\no & &
- \overline{u_L} \left( \frac{\varphi_1^+}{v_1} A_1
+ \frac{\varphi_2^+}{v_2} A_2 + \frac{\varphi_3^+}{v_3} A_3 \right) M_d d_R
\no & &
+ \overline{d_L} \left( \frac{\varphi_2^-}{v_2} A_1^\dagger
+ \frac{\varphi_1^-}{v_1} A_2^\dagger
+ \frac{\varphi_3^-}{v_3} A_3^\dagger \right) M_u u_R
+ {\rm H.c.}.
\label{Yukawas 2}
\ea

One sees from Eqs.~(\ref{scalar mixing}) and (\ref{Yukawas 2})
that the scalars $N_a$ couple to the quarks
with an Hermitian matrix multiplied on the right by the quark mass matrix.
The pseudoscalars $A_1$ and $A_2$ have similar Yukawa couplings,
with the Hermitian matrix substituted by an anti-Hermitian matrix.
Under these conditions,
the self-energies $\Sigma_d$ and $\Sigma_u$
following from the diagram in Fig.~1 satisfy
\ba
\Sigma_d M_d^{-1} = \left( \Sigma_d M_d^{-1} \right)^\dagger,
\no
\Sigma_u M_u^{-1} = \left( \Sigma_u M_u^{-1} \right)^\dagger,
\label{hermitian Sigmas}
\ea
and do not contribute to $\theta_{\rm QFD}$.

Now consider the up-quark self-energy $\Sigma_u$
following from a loop with the charged scalar $H_1^+$,
depicted in Fig.~2.
\begin{figure}[tb]
\begin{center}
\leavevmode \epsfxsize=120mm \epsfbox{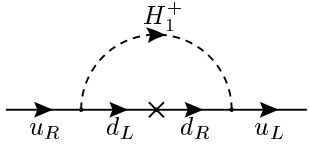}
\end{center}
\vspace*{-12mm}
\caption{One-loop diagram for $\Sigma_u$ with a charged scalar.}
\label{Georgi's diagram, 2}
\vspace*{-3mm}
\end{figure}
Let us write $H_1^+ = \sum_{a=1}^3 c_a \varphi_a^+$,
where the coefficients $c_a$ are {\em real}.
Then,
Fig.~2 yields
\be
\Sigma_u M_u^{-1} \sim \left(
\frac{c_1}{v_1} A_1 + \frac{c_2}{v_2} A_2 + \frac{c_3}{v_3} A_3
\right) f \left( M_d^2, m_{H_1}^2 \right) \left(
\frac{c_2}{v_2} A_1^\dagger + \frac{c_1}{v_1} A_2^\dagger
+ \frac{c_3}{v_3} A_3^\dagger \right),
\ee
where $f$ is the real function issuing from the loop integration.
Using the cyclic property of the trace
together with Eq.~(\ref{A A = delta}),
one finds that $\mbox{tr} \left( \Sigma_u M_u^{-1} \right)$ is real,
therefore it does not contribute to $\theta_{\rm QFD}$.

Thus,
$\theta_{\rm QFD}$ vanishes at one-loop level.
It is important to emphasize the crucial role that real CP violation
plays in attenuating the strong CP problem in this model.
Indeed,
it is this peculiar form of SCPV which allows CP violation
to be absent from the scalar mixing,
even while it persists in the fermion sector.
This means that {\em the $\rho_a$ do not mix with the $\eta_a$},
and also that the physical charged scalars
are {\em real} linear combinations of the $\varphi_a^+$.
The same thing happened in the previous model of real CP violation
that I have put forward \cite{lavoura}.

Unfortunately,
in the present model $\theta_{\rm QFD}$ arises at two-loop level,
and it is not clear to me that there is any suppression mechanism
which can make it small enough.
Therefore,
I cannot claim that this model completely avoids the strong CP problem.
To be fair,
however,
I must emphasize that there are very few viable models
which solve the strong CP problem
in the context of extensions of the standard model only with extra scalars;
in particular,
invisible-axion models \cite{invisible},
are tightly constrained by experimental tests \cite{axions}.

It must also be pointed out that this model
has unsuppressed flavour-changing neutral Yukawa interactions,
which may give undesirably large contributions to CP violation,
$K^0$--$\overline{K^0}$ mixing,
$B^0_d$--$\overline{B^0_d}$ mixing,
and the decay $K_L \rightarrow \mu^+ \mu^-$.

\section{Concluding remarks}

Spontaneous CP violation may provide a way of eliminating
the strong CP problem of the standard model \cite{parada}.
In order to reach that goal,
however,
it is usually important that CP violation remains absent
from the scalar mixing.
This may be achieved if SCPV does not hinge on the generation
of non-trivial phases among the VEVs,
but rather on the generation of real
(or else with trivial relative phases)
VEVs,
which however do not satisfy the equalities among themselves
implied by the initial CP symmetry.
We may call this situation ``real CP violation''.

As far as I can see,
the simplest model displaying these features involves,
for an odd number of fermion generations,
three Higgs doublets.
However,
if there were four (or, more generally, an even number of) generations,
two Higgs doublets would be enough.
The CP transformation would read
\be
\phi_1 \stackrel{\rm CP}{\rightarrow} {\phi_2^\dagger}^T,\
\phi_2 \stackrel{\rm CP}{\rightarrow} {\phi_1^\dagger}^T,\
\ee
\be
q_{L1} \stackrel{\rm CP}{\rightarrow} \gamma^0 C \overline{q_{L3}}^T,\
q_{L2} \stackrel{\rm CP}{\rightarrow} \gamma^0 C \overline{q_{L4}}^T,\
q_{L3} \stackrel{\rm CP}{\rightarrow} \gamma^0 C \overline{q_{L1}}^T,\
q_{L4} \stackrel{\rm CP}{\rightarrow} \gamma^0 C \overline{q_{L2}}^T,
\ee
and an extra discrete symmetry would change the signs of $\phi_2$,
$q_{L3}$,
and $q_{L4}$.
One would thus obtain fermion mass matrices of the form
\be
M_n = \left( \begin{array}{c} v_1 e^{i \theta_1} \Gamma \\
v_2 e^{i \theta_2} \Gamma^\ast \end{array} \right),\
M_p = \left( \begin{array}{c} v_1 e^{- i \theta_1} \Delta \\
v_2 e^{- i \theta_2} \Delta^\ast \end{array} \right),
\ee
with $\Gamma$ and $\Delta$ being $2 \times 4$ matrices.
CP violation in fermion mixing would be present as long as $v_1 \neq v_2$,
yet $\theta_{\rm QFD}$ would vanish both at tree level
and one-loop level.\footnote{The scalar potential of such a model
would need either soft CP breaking or extra scalar fields
in order to produce $v_1 \neq v_2$.}

More generally,
I have shown in this paper that real CP violation
is a distinct possibility whenever the Lagrangian respects
a non-trivial CP symmetry together with some other discrete symmetry.
Real CP violation may constitute an useful option for spontaneous CP breaking
for other purposes besides solving the strong CP problem;
is particular,
the non-trivial CP transformation associated with real CP violation
may be used to obtain specific relationships
among the fermion masses and mixings.

\end{document}